\documentstyle[12pt,aasms4]{article}

\lefthead{Roscherr \& Schaefer} \righthead{Light Echoes and IIn
Supernovae}

\begin{document}

\title{Can Light Echoes Account for the Slow Decay of Type IIn
Supernovae?}

\author{Bruce Roscherr and Bradley E. Schaefer} \affil{Physics
Department, Yale University, PO Box 208120, New Haven, CT, 06520-8121;
bruce.roscherr@yale.edu, schaefer@grb2.physics.yale.edu}

\begin{abstract}
The spectra of type IIn supernovae indicate the presence of
apre-existing slow, dense circumstellar wind (CSW). If the CSW extends
sufficiently far from the progenitor star, then dust formation
shouldoccur in the wind. The light from the supernova explosion will
scatter off this dust and produce a light echo. Continuum emission
seen after the peak will have contributions from both this echo as
well as from the shock of the ejecta colliding with the CSW, with a
fundamental question of which source dominates the continuum. We
calculate the brightness of the light echo as a function of time for a
range of dust shell geometries, and use our calculations to fit to the
light curves of SN 1988Z and SN 1997ab, the two slowest declining IIn
supernovae on record. We find that the light curves of both objects
can be reproduced by the echo model. However, their rate of decay from
peak, color at peak and their observed peak absolute magnitudes when
considered together are inconsistent with the echo model. Furthermore,
when the observed values of M$_{B}$ are corrected for the effects of
dust scattering, the values obtained imply that these supernovae have
unrealistically high luminosities. We conclude that light echoes
cannot properly account for the slow decline seen in some IIn's, and
that the shock interaction is likely to dominate the continuum
emission.
\end{abstract}

\section{Introduction}
Supernovae are designated to be of type IIn based on their spectra
(Schelegel 1990). The group is quite heterogeneous, but a typical IIn
would have the following properties: an $H\alpha$ profile composed of
a narrow peak ($\sim 100$ km/s) sitting on a broad base ($\sim 10^{4}$
km/s), no P Cygni absorption feature, a strong blue continuum, and
slow spectral evolution. There is occasionally an intermediate width
component ($\sim 2000$ km/s) to the $H\alpha$ line.  Absorption lines
tend to be weak or absent. The narrow $H\alpha$ component is seen to
vary in intensity and has thus been interpreted to indicate the
presence of a pre-existing slow-moving stellar wind. These winds can
be very dense, with $\dot{M}$ as high as $10^{-2}$ $M_{\odot}$/yr as
is the case with SN1997ab (Salamanca {\it et.al.} 1998).  The overall
spectral properties of these supernovae have successfully been
interpreted as being due to a shock interaction between the rapidly
expanding supernova ejecta and this dense ($10^{6}-10^{8}
\mbox{cm}^{-3}$) stellar wind (Chevalier \& Fransson 1985, Chugai
1991, Terlevich 1994). IIn's comprise about 5\% of observed type II
supernovae.  \par A number of IIn's show an abnormally slow decline in
their light curves.  SN1997ab and SN 1988Z are the extreme examples of
this, both declining by only about 5 magnitudes in 1000 days. Simple
models of the shock interaction suggest that the shock can produce
sufficient optical continuum emission to account for this slow decline
(Terlevich 1994, Plewa 1995). It was suggested by Chugai(1992) that
light echoes might also provide a significant contribution to the late
time light curves of IIn's, i.e. the observed continuum should be a
combination of shock and echo contributions. When the supernova
explodes, it vaporizes any dust in the CSW out to a radius of $\sim
10^{16}$ cm (Pearce \& Mayes 1986). Any dust that lies outside this
evaporation radius will scatter the supernova light. Scattered light
travels a longer path before reaching us and so is delayed relative to
any unscattered light, i.e.  a light echo is produced. As little as
$10^{-6} M_{\odot}$ of dust is required to make the circumstellar
shell of matter optically thick to dust scattering.  \par We here
investigate whether light echoes could help account for the slow
decline in the light curve seen in some IIn's. We have performed a
detailed calculation of the light echo produced by a range of dust
shell geometries. Our assumptions and method of calculation are
described in the next section. Our results are presented in Section 3,
and in Section 4 we fit to the observed light curves of SN 1988Z and
SN 1997ab. Section 5 is devoted to the discussion of our results.

\section{Calculating the Light Echo}
The light echo in the $U$, $B$, $V$, and $R$ bands was computed by a
Monte Carlo simulation. The calculation requires knowledge of the dust
size distribution, the dust spatial distribution, the optical
properties of the dust, and the underlying Type II supernova light
curve.
\par The dust was assumed to be composed of a mixture of
silicon and graphite grains, in the ratio 3:1 by number. Both grain
types were assumed to have a grain size distribution described by the
MRN grain model (e.g. Evans 1994), i.e. the number density of grains
varies as $n(a) \propto a^{-3.5}$ where $a$ is the grain diameter. Our
grain size ranged from $a_{min} = 0.005 \mu$m up to $a_{max} = 1.0
\mu$m.  \par We assumed that the dust lies in a spherical shell of
inner radius $R_{min}$ and outer radius $R_{max}$ about the
star. $R_{min}$ is the radius out to which the supernova evapourates
all the dust in the CSW. $R_{min}$ can be estimated from first
principles and for a type II supernova has a value of a few times
$10^{16}$ cm (Pearce \& Mayes 1986). We used this value as a guide,
and took both $R_{min}$ and $R_{max}$ to be free parameters. We
assumed that the dust density in the shell varies as $\rho (r) \propto
r^{-2}$. The amount of dust in the shell is characterized by $\tau$,
the $R$ band optical depth for a purely radial photon trajectory. The
total dust mass is related to $\tau$ by the following expression:
\begin{eqnarray}
 M_{dust} &=&
     \frac{8\pi}{3}\left(a_{max}^{0.5}-a_{min}^{0.5}\right)R_{max}R_{min}
     \left(f\rho_{gr}+(1-f)\rho_{Si}\right)\times \nonumber\\ & &
     \frac{\tau}{\int_{a_{min}}^{a_{max}} da\,
     a^{-1.5}\sum_{i=abs,scat}\left(f Q_{i,gr}+(1-f)
     Q_{i,Si}\right)}\nonumber\\ &=& 7.0\times10^{-7}
     \,\tau\left(\frac{R_{min}}{0.01\mbox{pc}}\right)\left(\frac{R_{max}}
{1.0\mbox{pc}}\right)
     \,\,M_{\odot}
\end{eqnarray}
where $f = 0.25$, $\rho_{gr} = 2.5 \mbox{g}\,\mbox{cm}^{-3}$ and
$\rho_{Si} = 2.35 \mbox{g}\,\mbox{cm}^{-3}$ are the densities of
graphite and silicon respectively, and $Q_{abs}$ and $Q_{scat}$ are
the absorption and scattering efficiencies.  
\par The photons are
either absorbed by the dust, or they scatter elastically.  The cross
sections for the dust-photon interactions are functions of both the
photon wavelength and the dust grain size. We used the scattering and
absorption cross sections, $\sigma_{abs}(=4\pi a^{2} Q_{abs})$ and
$\sigma_{scat} (= 4\pi a^{2}Q_{scat})$, calculated by
Draine(1987). The angular distribution for the scattering is given by
the Henyey-Greenstein function (Henyey \& Greenstein 1941),
\begin{equation}
 F(\theta) = \frac{1-g^{2}}{\left(1+g^{2}-2g\cos\theta\right)^{3/2}}
\end{equation}
where $g$ is the degree of forward scattering and is a function of
 wavelength and grain size. Values for $g$ were also taken from
 Draine(1987).  
\par We used the composite II-L and II-P light curves
 calculated by Doggett and Branch(1985) as input.  They provide
 averaged light curves for the $B$ and $V$ bands. We took the $U$ and
 $B$ bands to be initially identical, and likewise the $V$ and
 $R$. The input supernova was taken to have $B-V = 0$ at peak.
 \placefigure{fig1} 
\par Our Monte Carlo code follows the photon
 trajectories in three dimensions.  It keeps track of the time delay
 for each photon, as well as the photon weight. The weight was
 decreased at each interaction by the absorption probability. The
 photon was also split at each interaction into a part which has no
 further interactions and thus escapes, and a second part which is
 forced to interact at least once more. This substantially improves
 the output statistics. Our code was tested against a number of
 analytical test cases, and also against the calculations of
 Chevalier(1986), who calculated light echoes in the limit of low
 optical depth.  \par Our code produces the $R$ band light curve as
 well as the $V-R$, $B-V$, and $U-B$ colors as output. The output is
 dependent on the choice of $R_{min}$, $R_{max}$, and $\tau$.
 $10^{6}$ input photons were used to calculate each band for each
 choice of the model parameters. Our output does not include any
 contribution from line emission.

\section{Results}
Our echo model has three parameters: $R_{min}$, $R_{max}$ and
$\tau$. Figures 2,3, and 4 show the effect of varying each of these in
turn.  \placefigure{fig2} \placefigure{fig3} \placefigure{fig4} We see
from the figures that the echo is fairly insensitive to both $R_{min}$
and $R_{max}$ as long as the shell is thicker than a few hundredths of
a parsec. As the shell becomes thinner, the light curves fall off
faster, and more closely trace the intrinsic light curve. We can also
see that the presence of a small amount of dust has a pronounced
effect on the light curve at late times. The optical depths in the
$U$, $B$, $V$ and $R$ bands are always in the ratio
1.56:1.35:1.20:1.00. The $V-R$, $B-V$ and $U-B$ colors remain nearly
constant after about 200 days for all the non-thin shell cases. For
the the thin shells the colors tend to redden at late times.  \par
While there are variations amongst models, the following relations
serve as useful approximations:
\begin{eqnarray*}
(U-B)_{peak} \sim 0.17\tau\!\!&,&\!\!(B-V)_{peak} \sim (V-R)_{peak}
\sim 0.15\tau\\ \Delta m_{U}\sim 1.4\tau\,\,,\,\, \Delta m_{B}\sim
1.3\tau\!\!&,&\!\!\Delta m_{V}\sim 1.1\tau,\Delta m_{R}\sim 0.95\tau\\
\beta_{100,U}\sim 4.3-1.5\tau,\beta_{100,B}\sim 4.3-1.4\tau\!\!&,&\!\!
\beta_{100,V}\sim 4.0-1.2\tau,\beta_{100,R}\sim 4.0-1.1\tau
\end{eqnarray*}
$\Delta m$ is the number of magnitudes by which the peak of the light
curve has faded due to dust scattering compared to the peak in the
composite type II light curve, and $\beta_{100}$ is the number of
magnitudes by which the light curve drops from peak over the first 100
days after peak.

\section{Fits to SN 1988Z and SN 1997\mbox{ab}}
SN 1988Z and SN 1997ab are the two slowest declining IIn's on
record. (SN 1995N has a similar decay rate, but data on this object is
not yet available). The slow decline suggests that these objects are
embedded in a dense CSW. This is bourne out by analysis of the spectra
of these objects. Salamanca {\it et.al.}(1998) calculate that the
electron density in the CSW of SN 1997ab has a value close to $10^{8}$
cm$^{-3}$. Chugai(1992) found a similar result for SN 1988Z. We have
used our echo model to fit to the light curves of these two
supernovae.

\subsection{SN 1988Z}
SN 1988Z was discovered on 1988 December 12 in the galaxy
MGC+03-28-022 (Cappelloro \& Turatto, 1988). The supernova was past
peak at this time, but its high luminosity (m$_{B}$ = 16.75, M$_{B}$ =
-18.3 for $H_{0}= 65\, \mbox{kms}^{-1} \mbox{Mpc}^{-1}$) argues that
the maximum occured not long before discovery.  The supernova faded
monotonically at an unusually slow rate. Over the first 137 days it
faded by only 1.5 magnitudes in both the $B$ and $V$ bands, whereas
the composite II-L light curve drops by some 5 magnitudes over the
same period. The color evolution of SN 1988Z is very unusual (Turatto
{\it et.al.} 1993). Over the first 100 days $B-V$ decreases from 0.4
to 0.1 mag. The supernova then becomes redder, until the color reaches
a constant value of about 0.65 mag after 700 days.  \par Figure 5
shows our fit to the $B$ and $V$ band light curves of SN 1988Z.  The
$B$ band fit corresponds to $\tau = 2.0\pm 0.2$, $R_{min} =
0.10\pm0.02$ pc, $R_{max} = 0.35\pm 0.05$ pc. This corresponds to a
dust mass in the shell of (eq. (1)) $(4.9\pm1.1)\times10^{-6}
M_{\odot}$. The narrow component of the H$\alpha$ line is unresolved,
so if we adopt a value of 100 kms$^{-1}$ for the wind velocity, and
assume a constant mass loss rate, we infer that the progenitor star
emitted material for about 3400 years before going supernova at a rate
of $(1.3\pm0.3)\times 10^{-6}\,
\left(\frac{f_{dust}}{10^{-3}}\right)^{-1}\,
\mbox{M}_{\odot}\mbox{yr}^{-1}$, where $f_{dust}$ is the fraction of
the wind mass which has formed into dust. The fit to the $B$ band
determines the fit to the $V$ band, i.e. the $V$ band is not fit
separately. In both cases, the fit is reasonably good up to about 800
days, after which the observed light curves are seen to flatten
significantly with respect to the model predictions.
\placefigure{fig5}

\subsection{SN 1997ab}
SN 1997ab was discovered on 1996 April 11 in the galaxy HS 0948+2018
(Hagen \& Reimers 1997). It had a $B$ magnitude of 14.7 which
corresponds to an absolute magnitude, $M_{B}= -19.1$ ($H_{0} = 65
\mbox{kms}^{-1}\mbox{Mpc}^{-1}$). The date of peak is uncertain, but
the high luminosity at discovery argues that the peak had occured
recently. The light curve of this object is unfortunately not well
sampled, there being only three other $B$ band observations and two
$V$ band observation since discovery (Hagen, Engels \& Reimers 1997,
Schaefer \& Roscherr 1998, 1999).  \par Our model fits the $B$ band
light curve (figure 6) for $\tau = 3.0\pm 0.3$, $R_{min}= 0.10\pm
0.02$ pc, $R_{max} = 0.30\pm 0.05$ pc. This implies a dust mass in the
shell of $(6.3\pm1.4)\times 10^{-6} M_{\odot}$. The wind velocity is
measured to be 90 kms$^{-1}$, and so we infer that the progenitor
emitted material for some 3000 years before going supernova at an
average rate of $(2.9\pm0.7)\times 10^{-6}\,
\left(\frac{f_{dust}}{10^{-3}}\right)^{-1}\,\mbox{M}_{\odot}\mbox{yr}^{-1}$.
Salamanca et.al. (1998) find a mass loss rate of approximately
$10^{-2} M_{\odot}\mbox{yr}^{-1}$ from the analysis of the spectrum of
this object.  They find that the emitted material lies in a shell that
extends no more than $0.05$pc from the progenitor. This implies that
the progenitor had a brief episode of strong mass loss shortly before
going supernova. The material emited during this episode lies too
close to the star to allow any dust formed to survive the supernova
explosion. The dust which is responsible for the echo must have been
emitted during an earlier, more sustained period of lower mass loss.
\placefigure{fig6}

\section{Echo Model versus Shock Model}
If the progenitors of Type II supernovae emit material for more than a
few hundered years prior to going supernova, then they should be
surrounded by a dusty CSW. All the dust that lies beyond the
evapouration radius of the supernova will scatter the supernova light
and produce a light echo. There is likely to be an echo component in
the light curves of many Type II supernovae, the only quesiton is how
significant a component it is.  
\par We have seen in the previous
section that the echo model can provide reasonably good fits to the
light curves of SN 1988Z and SN 1997ab.  Chugai(1992), however,
produced a similar quality fit to the light curve of SN 1988Z with a
simple model of the shock continuum emission. Fits to the light curves
alone are thus not sufficient to discriminate between the models.
\par In figure 7 we have plotted$(B-V)_{peaK}$ against
$\beta_{100,B}$. The region between the two solid lines is the region
of parameter space allowed by the echo model. As the amount of
circumstellar dust increases, we expect the supernova light to suffer
more scatterings, producing an echo that becomes progressively redder
and longer lived. In shock models the optical light is primarily
produced by the reprocessing of X-rays. The color of the optical light
produced in this way is not very sensitive to the amount of
circumstellar material present, so on the plot of $(B-V)_{peak}$
versus $\beta_{100,B}$, we do not expect to see a correlation. SN
1988Z and SN 1997ab both lie within the region allowed by the echo
model. The other IIn's plotted on the figure, however, do not. Their
distribution shows no definite correlation, and is thus consistent
with the shock model prediction. The IIn's plotted represent the
complete sample for which $(B-V)_{peak}$, $\beta_{100,B}$, and
$M_{B,peak}$ are available in the literature (Patat {\it et.al.}
1994). The data for SN 1998S was taken from the CfA Supernova Group
website.  \par In figure 8 we have plotted the observed $M_{B,peak}$
versus $\beta_{100,B}$. In the echo model, for a given underlying
supernova luminosity, as the dust mass increases, the decay rate
decreases and the observed $M_{B,peak}$ increases, i.e. the supernova
appears fainter as more of the emission at peak is scattered by the
dust. The opposite trend is expected for shock models, i.e.  we expect
the peak luminosity to increase as the amount of circumstellar
material increases. In an $M_{B,peak}$ versus $\beta_{100,B}$ plot,
shock models thus predict a downward sloping line. The observed IIn
points are consistent with the shock prediction, and inconsistent with
the prediction of the echo model. In particular, the points for both
SN 1997ab and SN 1988Z lie outside the region allowed by the echo
model.  
\par In figure 9 we have plotted the observed $M_{B,peak}$
versus $(B-V)_{peak}$.  The echo model predicts that the emission
should redden as the amount of dust, and thus $M_{B,peak}$,
increases. The observed data points are inconsistent with this
prediction.  \par Another argument against the echo model is the
implied values for $M_{B,peak}$ for the underlying supernovae. SN
1988Z had an observed $M_{B,peak}$ of -18.3 and SN 1997ab, -19.1. If
we use the values of $\tau$, $R_{min}$ and $R_{max}$ obtained from the
fits to the light curves to find $\Delta m_{B}$, we find that the
underlying supernovae must have had $M_{B,peak}$ of -20.9 in the case
of SN 1988Z, and -22.8 for SN 1997ab. This would make these supernovae
by far the brightest Type II supernovae yet observed.  
\par We conclude that light echoes cannot properly account for the slow
decline seen in some IIn's. The predictions of shock models are
consistent with the data, and shocks are likely to dominate the
continuum emission. The absence of a strong echo component in the
light curves of IIn's argues that the progenitors of these supernovae
undergo the bulk of their mass loss just prior to going supernova, and
thus very little dust survives the explosion.

\clearpage
\figcaption[f1/ps]{(a) and (b) respectively show the composite II-L and the
composite II-P light curves used as input for the echo
calculations. (c) and (d) show the values of $B-V$ for each set of
composite curves Doggert \& Branch(1985).  \label{fig1}}

\figcaption[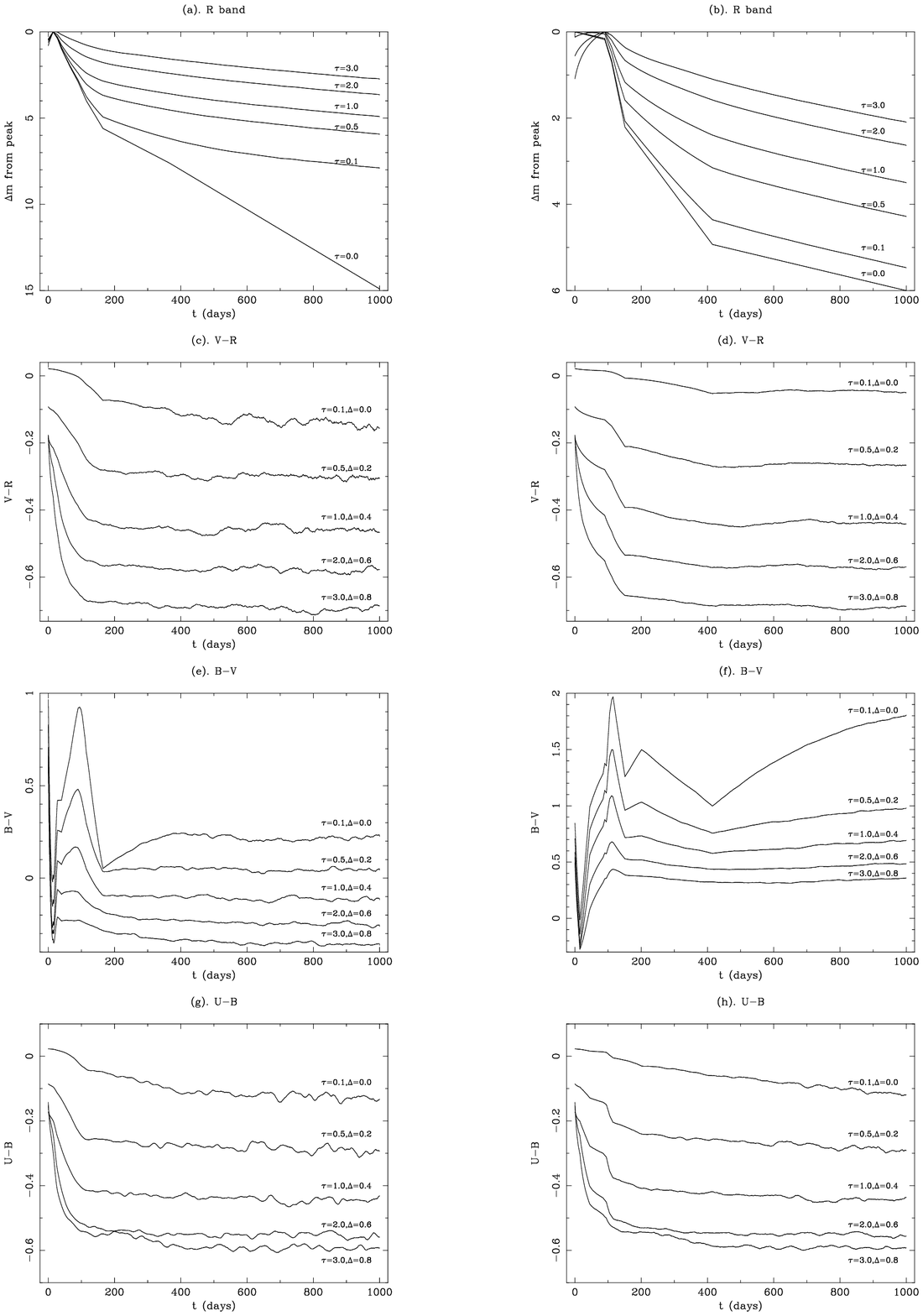]{The effect on the light curves due to changes in
$\tau$. R$_{min}$ and R$_{max}$ are held fixed at 0.1 and 1.0 pc
respectively. (a) and (b) show the $R$ band light curve for input II-L
and II-P respectively. The remaining plots are for $V-R$, $B-V$ and
$U-B$ for each of the two sets of input. The curves have been shifted
downward by $\Delta$ mags for clarity.\label{fig2}}

\figcaption[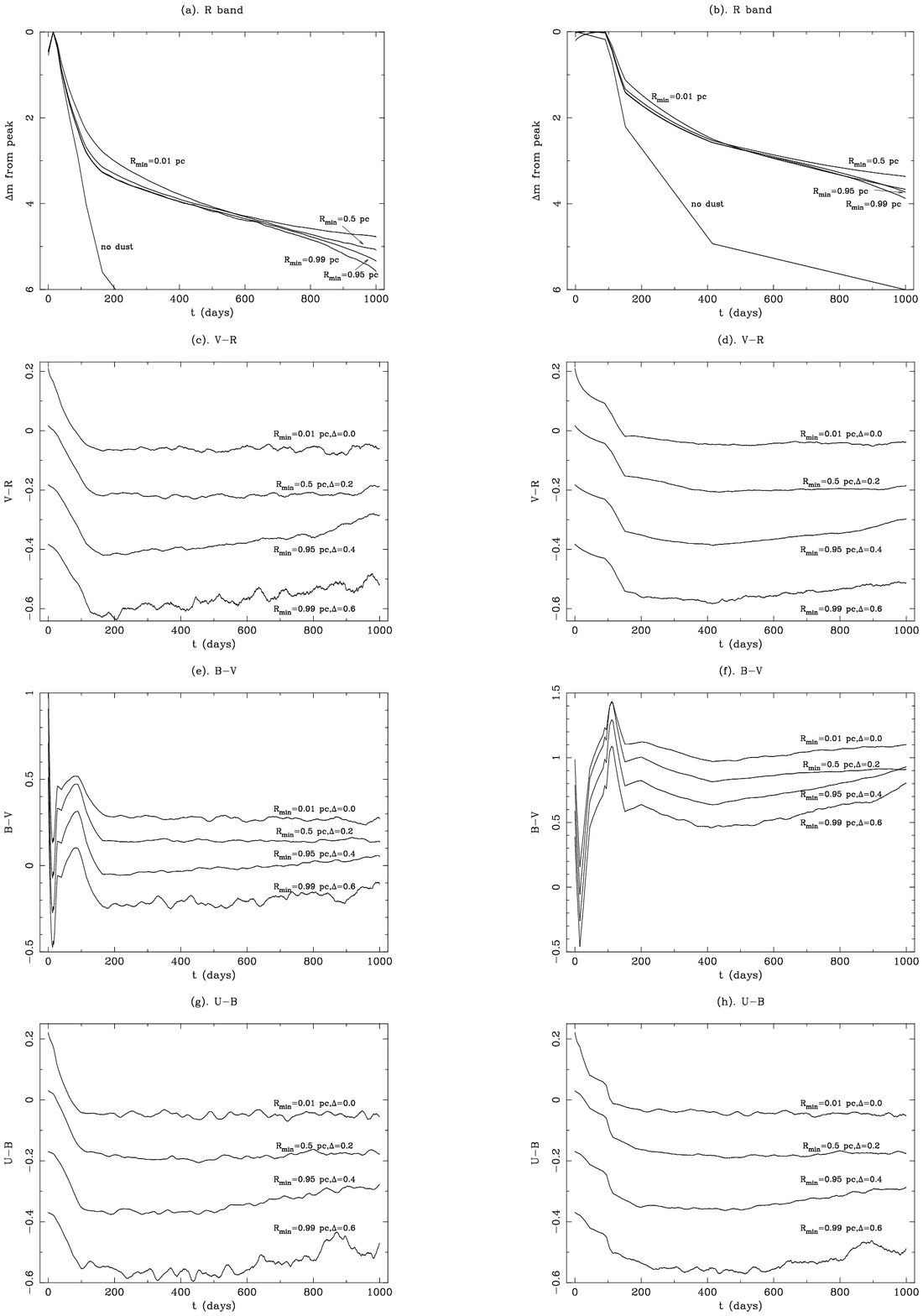]{The effect of changes in R$_{min}$. The curves are for
R$_{min}$ = 0.01, 0.5, 0.95 and 0.99 pc. $\tau$ is held fixed at 1.0
and R$_{max}$ at 1.0 pc. (a) and (b) show the $R$ band light curve for
input II-L and II-P respectively. The remaining plots are for $V-R$,
$B-V$ and $U-B$ for each of the two sets of input. The curves have
been shifted downward by $\Delta$ mags for clarity.\label{fig3}}

\figcaption[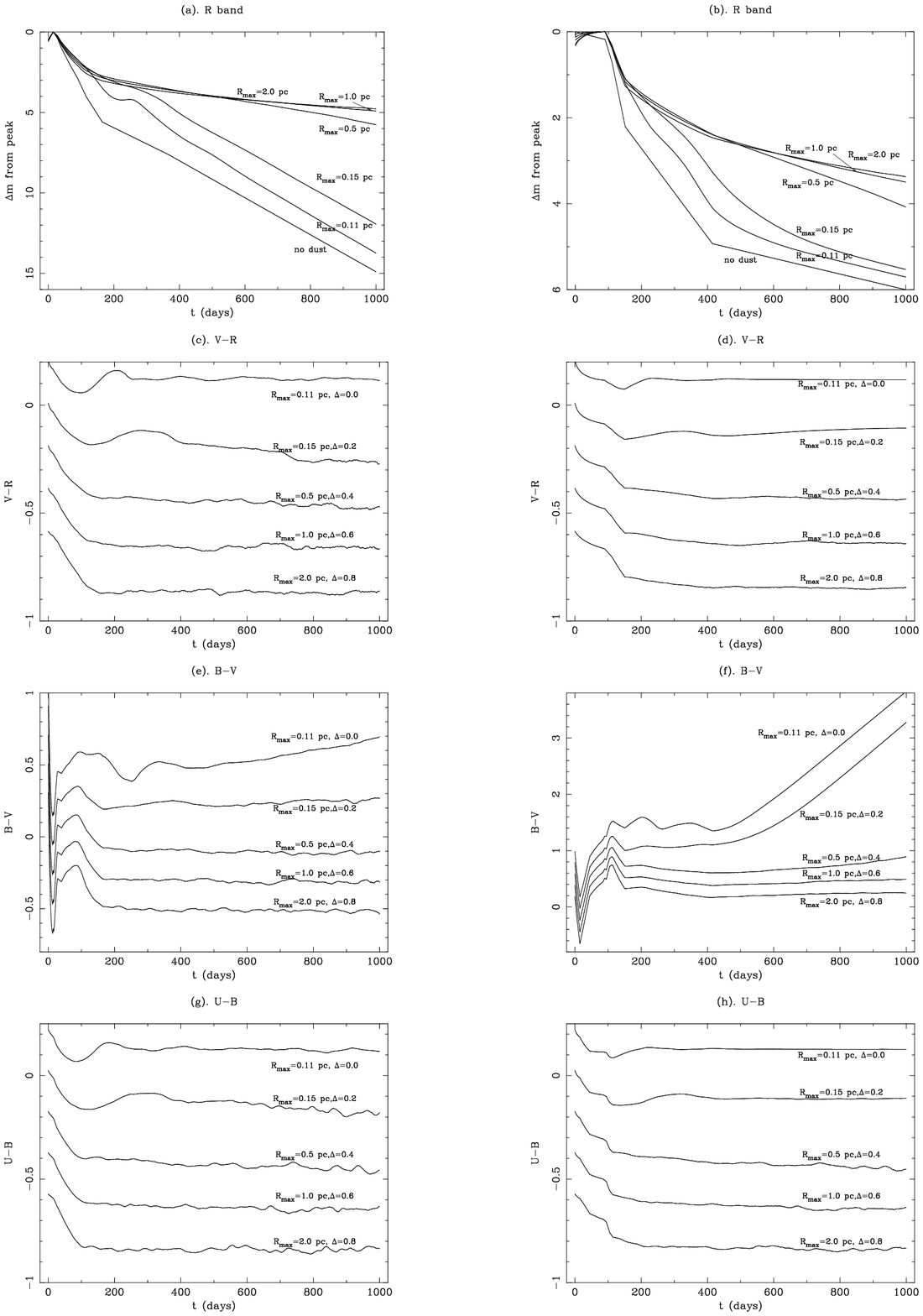]{The effect of changes in R$_{max}$. The curves are for
R$_{max}$ = 0.11, 0.15, 0.5, 1.0 and 2.0 pc. $\tau$ is held fixed at
1.0 and R$_{min}$ at 0.1 pc. (a) and (b) show the $R$ band light curve
for input II-L and II-P respectively. The remaining plots are for
$V-R$, $B-V$ and $U-B$ for each of the two sets of input. The curves
have been shifted downward by $\Delta$ mags for clarity.
\label{fig4}}

\figcaption[f5.ps]{Fit to the $B$ (crosses) and $V$ (circles) band light
curves of SN 1988Z (Turatto {\it et.al.} 1993).  \label{fig5}}

\figcaption[f6.ps]{Fit to the $B$ (crosses) and $V$ (circles) band light
curves of SN 1997ab. \label{fig6}}

\figcaption[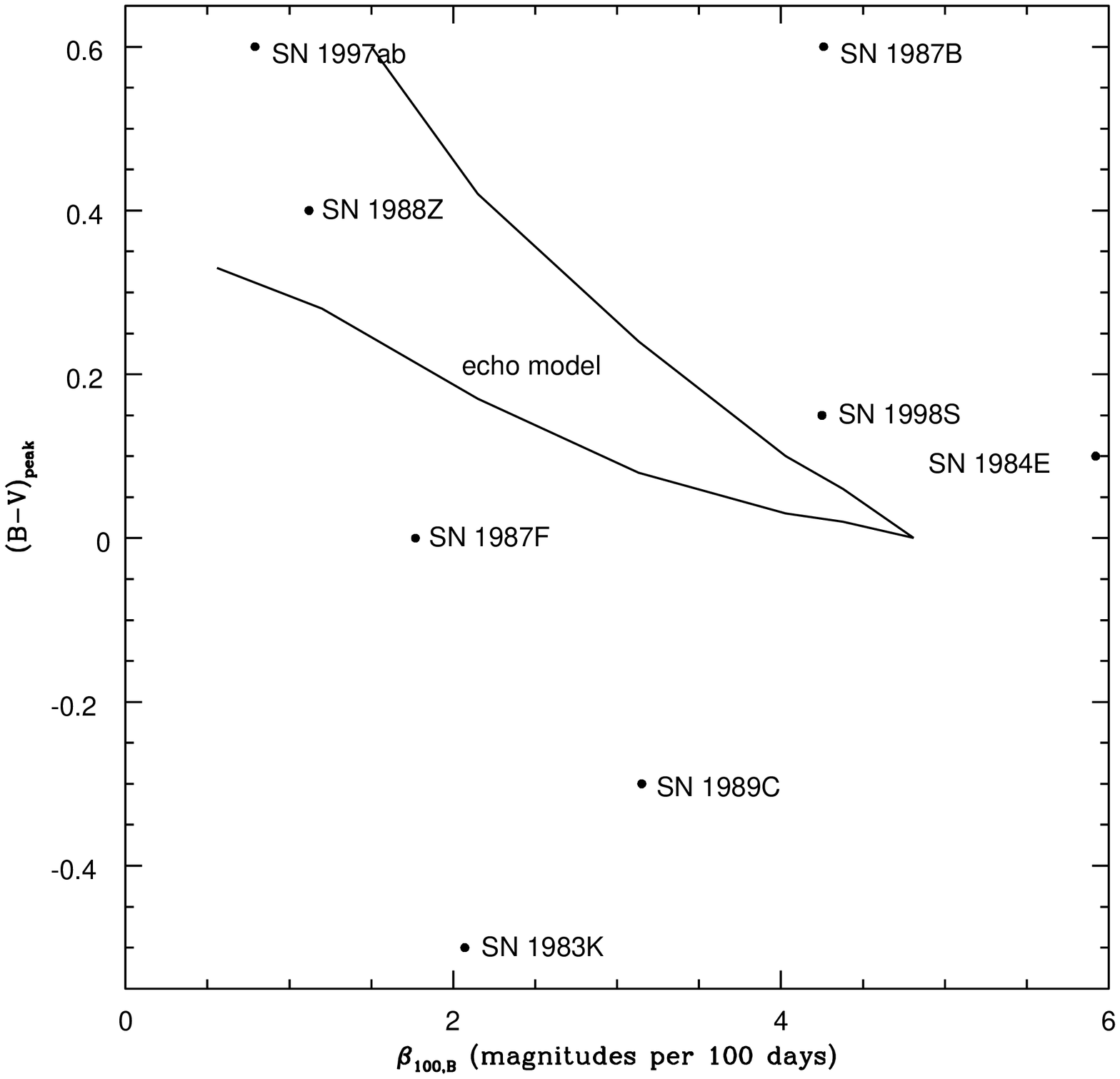]{Plot of $(B-V)_{peak}$ versus $\beta_{100,B}$. The area
between the solid lines are the values allowed by the echo model. The
underlying supernova was taken to have $B-V=0$ at peak.\label{fig7}}

\figcaption[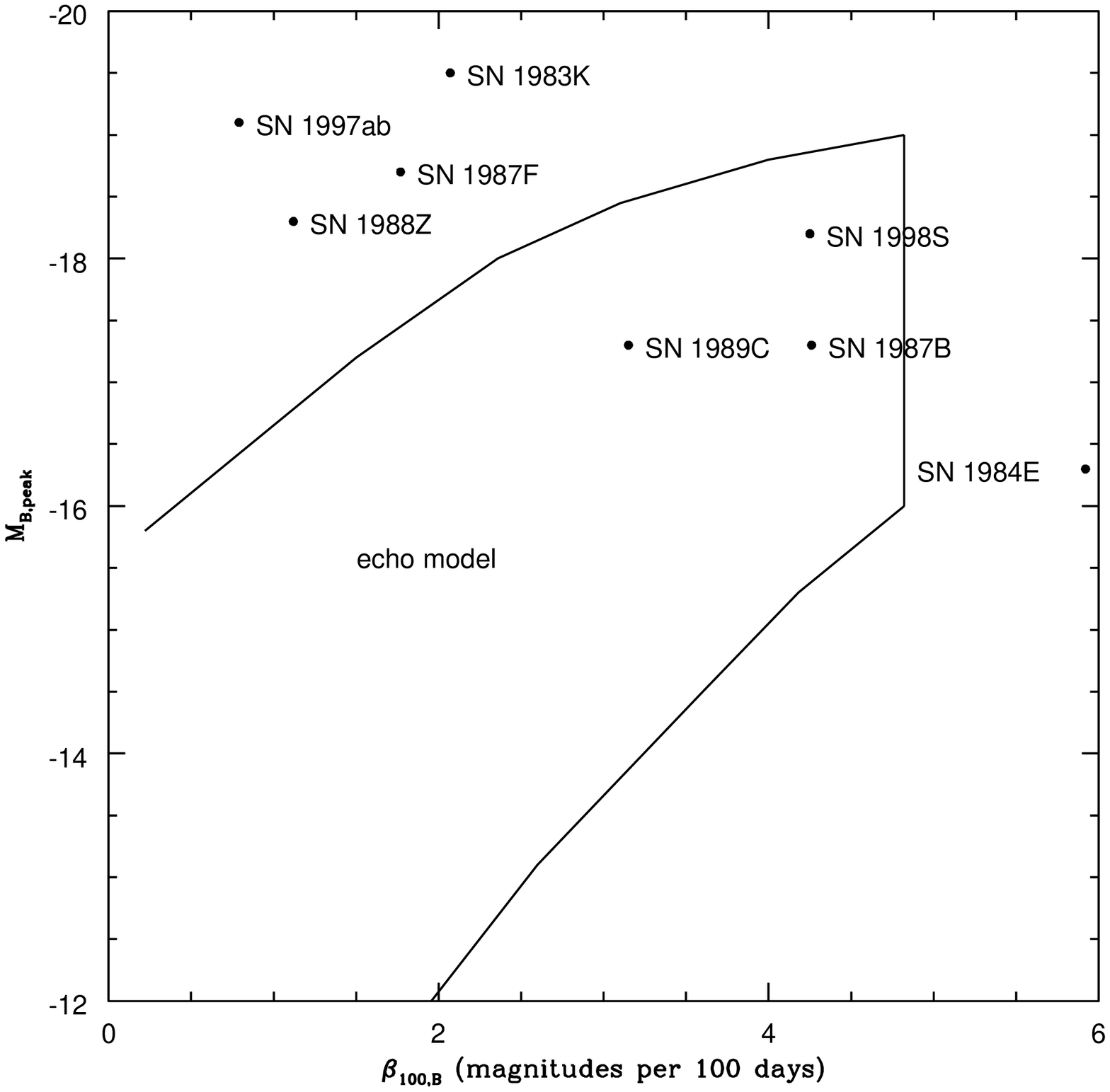]{Plot of $M_{B,peak}$ versus $\beta_{100,B}$. $M_{B,peak}$
for the underlying supernova was allowed to range between -16 and
-19.\label{fig8}}

\figcaption[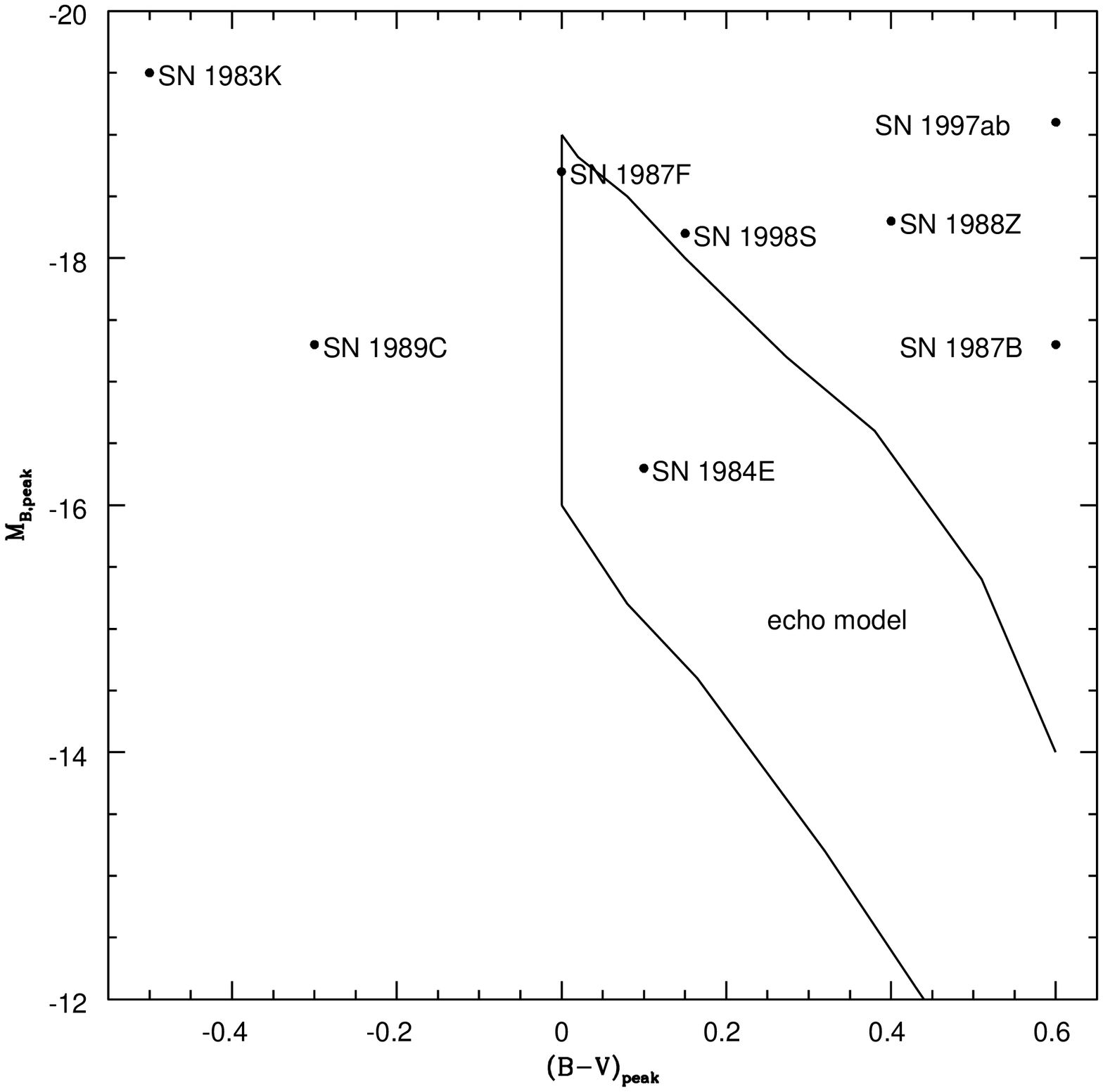]{Plot of $M_{B,peak}$ versus $(B-V)_{peak}$. For the
 underlying supernova, $M_{B,peak}$ was allowed to range between -16
 and -19, and we assumed $(B-V)_{peak} = 0$.  \label{fig9}}

\end{document}